\documentclass[aip,superscriptaddress,reprint]{revtex4-1}

\usepackage{graphicx}% Include figure files
\usepackage{dcolumn}% Align table columns on decimal point
\usepackage{bm}% bold math
\usepackage{color}
\usepackage{mathrsfs}
\usepackage{soul}

\begin{document}

%\preprint{AIP/123-QED}

\title{Voltage--Induced Buckling of Dielectric Films using Fluid Electrodes}

\author{Behrouz Tavakol}
\affiliation{Biomedical Engineering and Mechanics, Virginia Tech, Blacksburg, VA 24061.}
\affiliation{Mechanical Engineering, Boston University, Boston, MA 02215.}
\author{Douglas P. Holmes}%
\email{dpholmes@bu.edu.} 
\affiliation{Mechanical Engineering, Boston University, Boston, MA 02215.}

\date{\today}

\begin{abstract}
Accurate and integrable control of different flows within microfluidic channels is crucial to further development of lab-on-a-chip and fully integrated adaptable structures. Here we introduce a flexible microactuator that buckles at a high deformation rate and alters the downstream fluid flow. The microactuator consists of a confined, thin, dielectric film that buckles into the microfluidic channel when exposed to voltage supplied through conductive fluid electrodes. We estimate the critical buckling voltage, and characterize the buckled shape of the actuator. Finally, we investigate the effects of frequency, flow rate, and the pressure differences on the behavior of the buckling structure and the resulting fluid flow.  These results demonstrate that the voltage--induced buckling of embedded microstructures using fluid electrodes provides a means for high speed attenuation of microfluidic flow.
\end{abstract}

\maketitle

Dielectric actuators consist of an electrically insulating film that can be polarized through an applied electric field, commonly implemented by adhering the film between two electrodes.\cite{Pelrine1998} Upon exposure to an electric field, the opposing charges draw the electrodes closer together, developing a Maxwell stress within the film which reduces its thickness. The incompressible nature of the film causes its surface area to increase when its edges are free.\cite{Pelrine2000, Bozlar2012} Constraining the film from expanding by clamping its edges causes an additional compressive stress to develop within the film, which will cause it to buckle if it exceeds a critical magnitude determined by its geometry and material properties.\cite{Tavakol2014}  For large deformations, compliance is desired, and a variety of electrode systems have been utilized~\cite{Brochu2010} including carbon grease,\cite{Pelrine2000} silicone rubber embedded with carbon black,\cite{Bozlar2012} ionic elastomers,\cite{Keplinger2013} and electrodeless ion-bombardment.\cite{Rosset2009} In this Letter, we utilize a fluid electrode to buckle a thin film  embedded within a microfluidic channel.  The dielectric elastomeric film is surrounded by a conductive fluid, and actuated by passing an electric signal through the fluid which causes the clamped film to buckle out of the plane. We demonstrate the buckled film's ability to alter the fluid flow while being actuated at high frequencies and relatively low voltages. As controlling and directing fluid flow with microfluidic systems is crucial to enhancing their functionality,\cite{Whitesides2006, Mark2010, Whitesides2011} the development of microvalves and micropumps is essential,\cite{Unger2000,Melin2007} and our results demonstrate the feasibility of using an electric signal to actuate internally embedded structures.

\begin{figure}[h!]
\begin{center}
\vspace{2mm}
\resizebox{1\columnwidth}{!} {\includegraphics {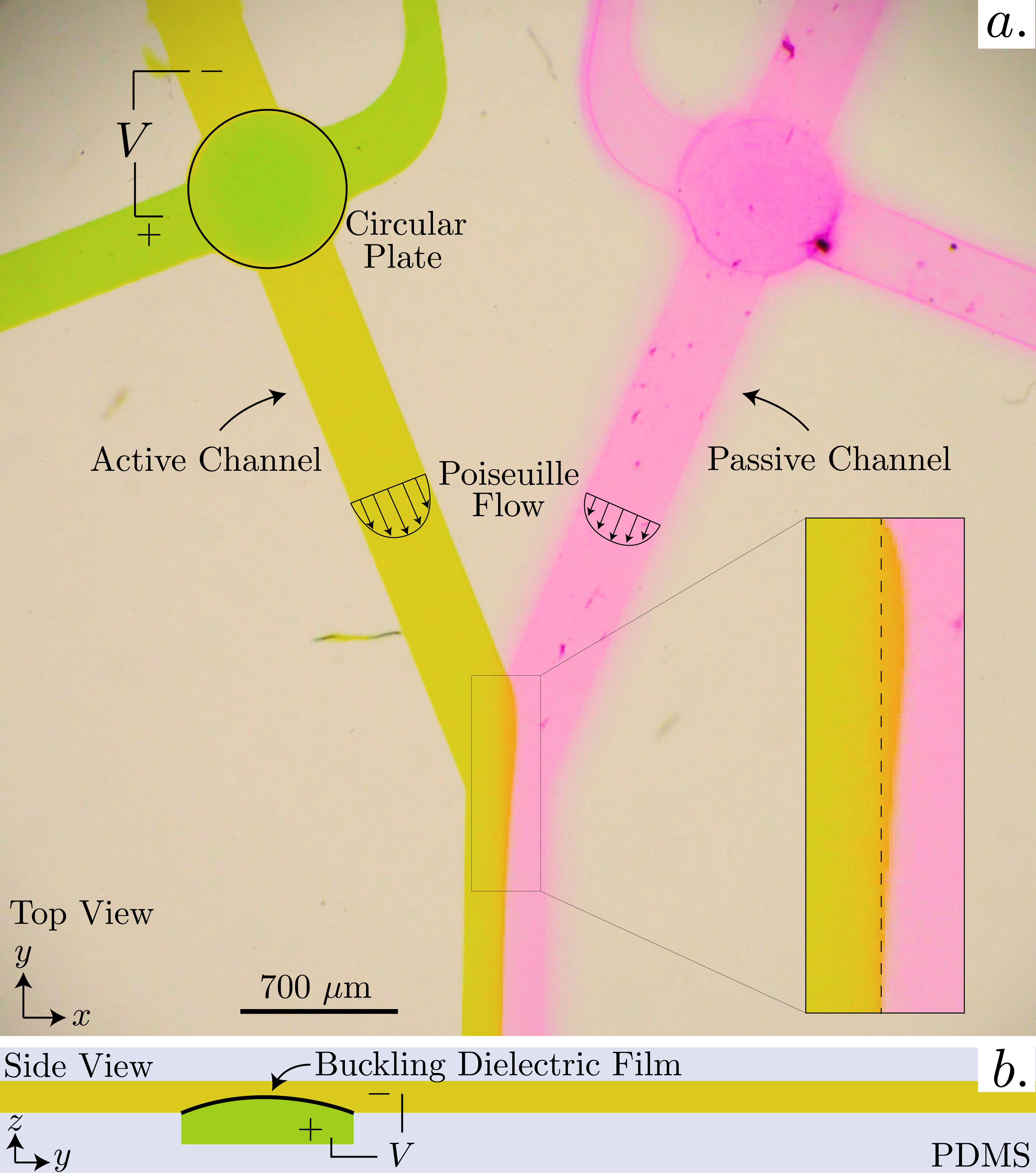}}
\end{center}
\vspace{-3mm}
\caption[]{{\sl a.} Image and top view schematic of a Y--junction multilayer microchannel with fluid electrodes and the buckling valve. {\sl b.} Side view schematic of a circular dielectric plate buckled into a spherical cap from an applied voltage. Positive voltage is applied to the fluid electrode below the dielectric plate, while the active microfluidic channel is grounded. \vspace{-6mm}
\label{fig1}}
\end{figure}

\begin{figure*} %[b!]
\begin{center}
\resizebox{1\textwidth}{!} {\includegraphics{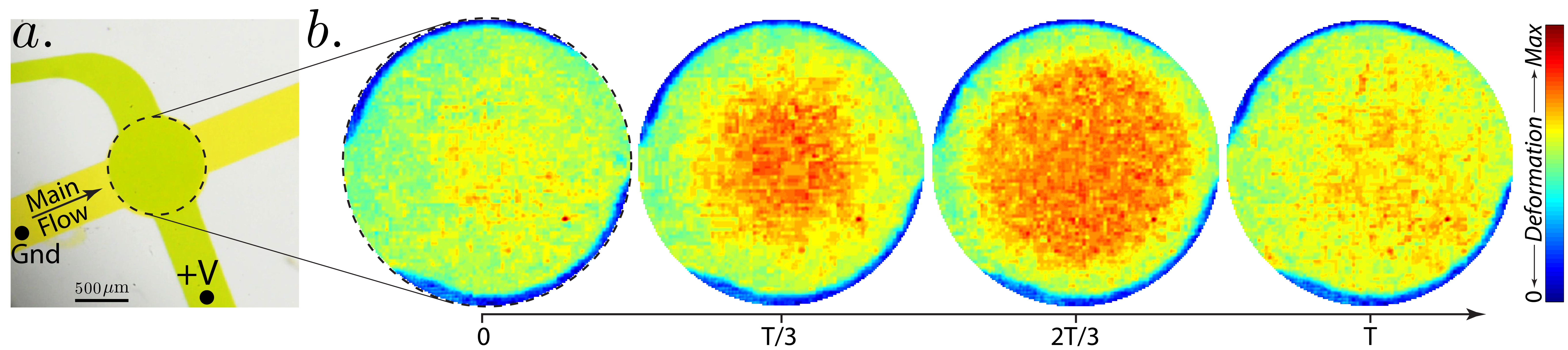}}
  \caption{ {\sl a.} Top view of the device showing the microactuator at the circular intersection of the active and the control channels. {\sl b.} 3D patterns of buckling deformations in one cycle. 
\label{fig2} \vspace{-5mm}}
\end{center}
\vspace{-2mm}
\end{figure*}

We demonstrate the fluid electrode microactuators within a microfluidic device where a Y-junction brings together two miscible fluid streams driven by an equivalent pressure difference $\Delta P$. Upstream of the Y-junction, we define an ``active'' channel which will be exposed to the electric field, and a ``passive'' channel which will only be subjected to the pressure--driven flow (Fig.~\ref{fig1}a). The fluid--structure interaction between the voltage--induced buckling of a dielectric plate and its surrounding microfluidic environment is achieved by preparing two vertically offset fluid channels separated by a thin film (Fig.~\ref{fig1}b). The fluid channels above and below the dielectric film are filled with an electrically conductive fluid, {\em e.g.} salt water, therefore, by grounding the upper channel and applying a positive voltage to the lower fluid we induce a Maxwell stress in the dielectric film. If this stress exceeds the critical buckling threshold of the plate, it will buckle out of the plane and constrict the microchannel. To bias the buckling in the direction that will form a constriction in the active channel, we apply a small amount of positive pressure to the channel below the dielectric film.

To experimentally determine the buckling behavior of the plate, we image the circular intersection between the upper and lower channels from above (Fig.~\ref{fig2}a), and use an image processing technique to estimate the shape of the buckled thin film. Applying voltage beyond the critical threshold causes the circular plate to buckle out of the plane, constricting the channel. Since the intensity of the fluid color depends on the fluid height at each point, we can determine the buckling shape of the plate by tracking the changes in the intensity of the dyed fluid. This technique is sensitive to small changes in intensity, thus enabling us to obtain a 3D deformation pattern. Fig.~\ref{fig2}b shows that the thin film is relatively flat at the beginning of a cycle, when the voltage is off. As the voltage increases, the plate buckles with an axisymmetric mode whose amplitude reaches its maximum at about 2/3 of the cycle $T$, followed by a quick return to its initial shape when the voltage goes back to zero. By tracking both the film deflection and the alteration of the downstream fluid flow, we observe a critical buckling voltage of $V_c = (3.3\pm0.2) \text{ kV}$.

To estimate the onset of buckling, we consider the stress generated in a dielectric material from an externally applied voltage in the manner presented by Pelrine {\em et al.}~\cite{Pelrine1998} The electrostatic energy stored in film is given by $\mathcal{U}=\frac{1}{2}Q^2/C$, where $Q$ is the charge and $C$ is the capacitance given by $C=\epsilon_0\epsilon A/h$ in terms of the free space permittivity $\epsilon_0$, the relative dielectric constant of the material $\epsilon$, and the ratio of the film's area to thickness, $A/h$. The pressure due to a change in electrostatic energy is $p_V=A^{-1}\mathcal{U}_{,h}$, where the comma denotes differentiation. As noted by Pelrine {\em et al.},~\cite{Pelrine1998} the essential difference between parallel--plate capacitors and deformable electrodes is the compliance of material, which results in doubling the stress in the film. This doubling arises from the stored electrostatic energy changing both the film thickness and area. Since the film is incompressible, {\em i.e.} $\nu\approx 0.5$, the film can deform to spread out similar charges on the film surface, and compress to bring opposing charges together, and these deformations will necessarily be coupled to the incompressibility constraint. An electric field is given by $\mathscr{E}=Q\left(\epsilon_0\epsilon A\right)^{-1}$, and can be approximated as $\mathscr{E}=V/h$ when the field is uniform. The resulting stress from the applied voltage is $p_V=- \epsilon_0\epsilon(V/h)^2$. From the incompressibility condition, it is apparent that for small strains the radial strain is half the transverse strain,\cite{Pelrine1998} {\em i.e.} $\varepsilon_r \approx -\frac{1}{2}\varepsilon_z$. The same must be true for the radial stress as the plate is clamped along its edge and constrained from expanding in plane. Therefore, we can determine the critical buckling voltage by comparing the radial stress imparted by the electric field, {\em i.e.} $\sigma_r=\frac{1}{2}p_v=\frac{1}{2}\epsilon_0\epsilon(V/h)^2$, with the stress required to buckle a circular plate. The critical buckling stress for a circular plate is determined by considering the linearized plate equations in cylindrical coordinates. For the plate geometry studied here, these equations have periodic solutions in the form of Bessel functions of the first kind, {\em i.e.} $J_1(k)$, and predict the critical buckling stress to be $\sigma_c=k^2B/a^2h$, where $B=\frac{Eh^3}{12(1-\nu^2)}$ is the bending energy, and $a$ is the plate radius~\cite{Timoshenko1961}. Equating the radial stress induced by the electric field with this critical stress -- while considering the plane stress condition -- results in the following equation for the critical voltage to buckle the plate:
\begin{equation}
V_c = \frac{kh^2}{a}\left[\frac{E}{6\,\epsilon_0\epsilon}\right]^{\frac{1}{2}},
 \label{eq_vc}
\end{equation}
where the constant for a clamped plate exhibiting the first buckling mode is $k^2=14.68$.
For example, here are the measured parameters for one of our samples: $h = 60\,\mu \text{m}, a=400\,\mu \text{m}, \epsilon \approx 3, \epsilon_0 = 8.85\times10^{-12}\,\text{N}/\text{V}^2, \text{and } E = 1.2 \text{ MPa}$. Equation \ref{eq_vc} therefore estimates that $V_c = 3.0 \text{ kV}$, which is in agreement with the experimentally measured critical voltage of $V_c = (3.3\pm0.2) \text{ kV}$.

\begin{figure}[t]
\begin{center}
\resizebox{1\columnwidth}{!} {\includegraphics{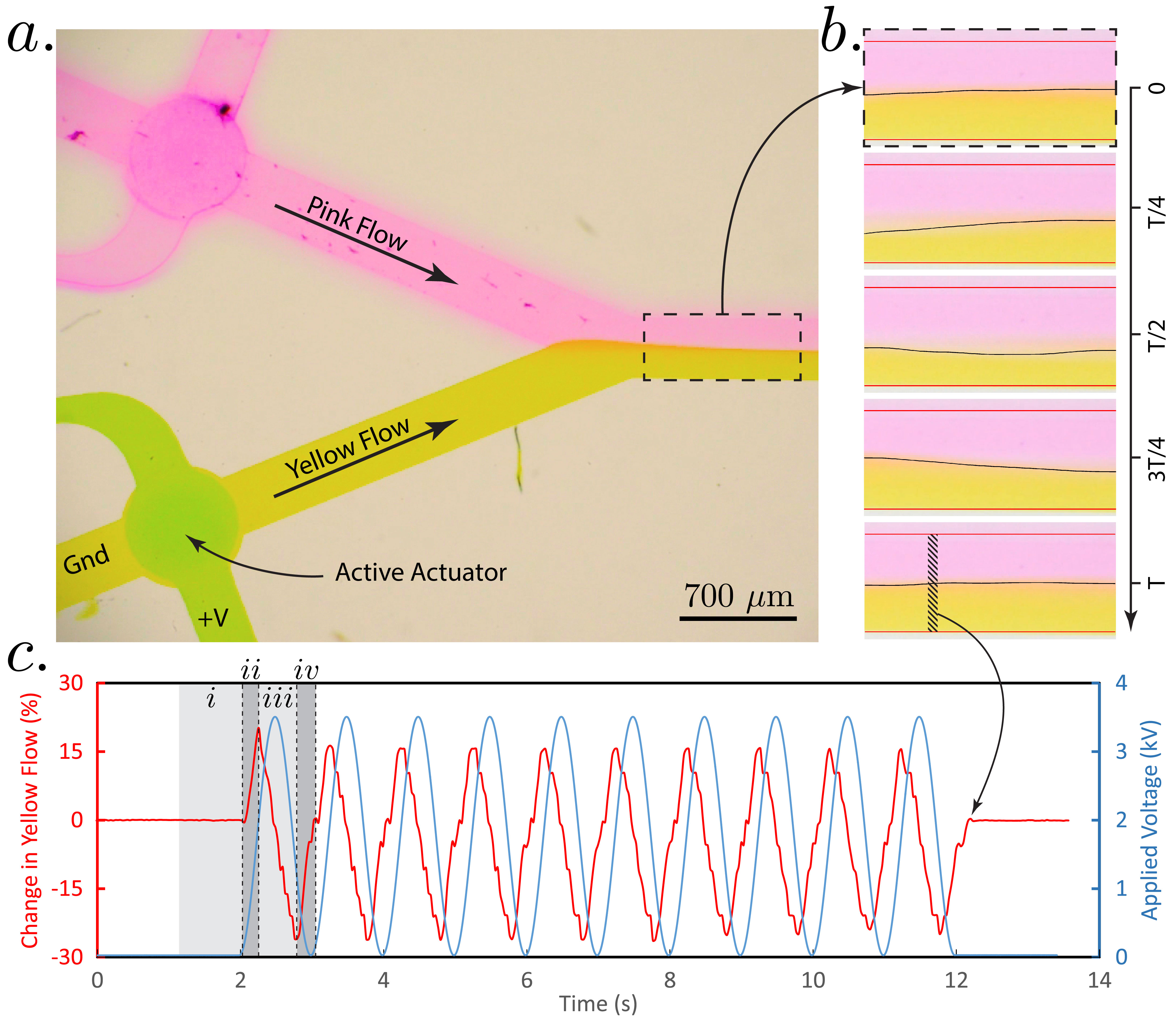}}\vspace{-2mm}
  \caption{
{\sl a.} Two pressure driven flows merging into one channel with a high voltage applied. {\sl b.} Sequences of images showing the disturbance of the fluid interface caused by the microactuator. The interface and the channel walls are overlaid on top of the raw images. {\sl c.} Changes in the ratio of yellow flow when the voltage is oscillating at a rate of 1 Hz. $i-iv$ bands demonstrate four different stages of coupling between the thin film's deformation and changes in fluid flow.
\label{fig3}} \vspace{-5mm}
\end{center}
\vspace{-2mm}
\end{figure}

We next study the interaction of the buckling plate with the adjacent fluid flow, which is evidenced by the the changes in ratio of one fluid's width relative to the channel width (Fig.~\ref{fig3}a). When the plate buckles out of the plane, it alters the flow in the corresponding channel, which consequently changes the ratio of the two fluids downstream of the Y-junction -- moving the interface up or down accordingly. We obtain the interface between two fluids by image processing (Fig.~\ref{fig3}b), and calculate the changes in the ratio of one flow by averaging the ratio of one flow across a narrow length shown in Fig.~\ref{fig3}b. \footnote{We ran each experiment for several cycles to improve the consistency of results (Fig.~\ref{fig3}c). In addition, we let the film relaxed for one second before switching to the next frequency test. This relaxation time is at least 3 orders of magnitudes longer than the time required for charging and discharging the micro actuator.} The interaction between the thin film and the adjacent fluid has four major steps during one voltage cycle. The first one is an equilibrium state where the voltage is zero (Fig.~\ref{fig3}c-i). By increasing the voltage, the thin film buckles into the channel, causing a momentary increase in the flow of that channel as fluid above the film is displaced (Fig.~\ref{fig3}c-ii). Next, the reduction in the pressure driven flow due to the buckled film's constriction of the active channel decreases the proportion of that fluid across the channel's width below the Y-junction (Fig.~\ref{fig3}c-iii). Finally, the voltage is reduced to zero and the film tends to go back to its original shape, bringing the flow downstream back to its original state. By repeating the voltage cycle, we observe the same interaction behavior, although the maximum change in flow becomes more consistent usually after the second cycle (Fig.~\ref{fig3}c).

\begin{figure}%[h!]
\vspace{0mm}
\begin{center}
\resizebox{1\columnwidth}{!} {\includegraphics{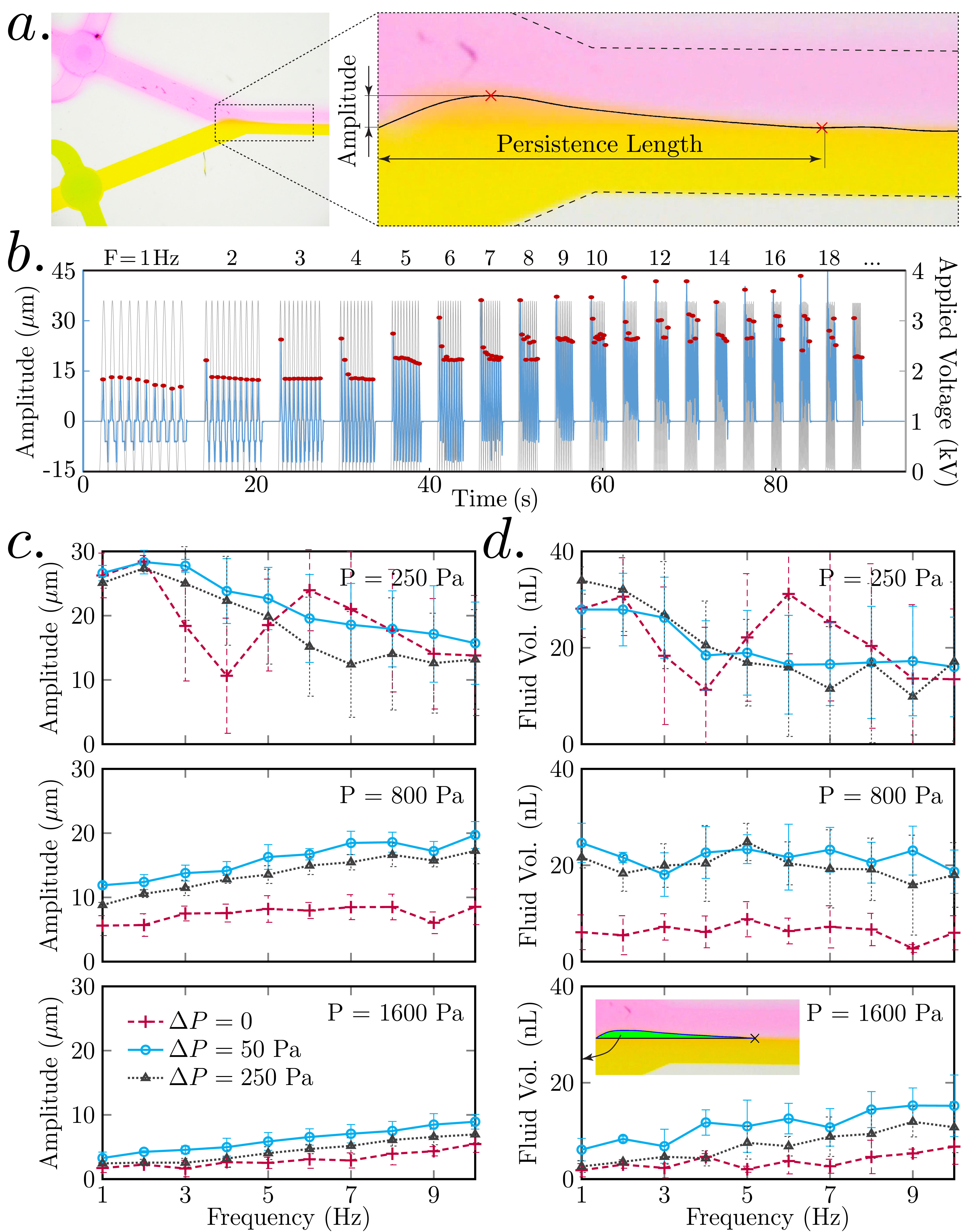}}\vspace{-1mm}
  \caption{
{\sl a.} An image of the fluid flow altered by the buckled microactuator. {\sl b.} A plot of the amplitude of the disturbed interface. Maximum amplitudes per each voltage cycle are shown in red dots. {\sl c.-d.} Changes in the maximum amplitude and the volume of disturbed fluid as a function of frequency for three pressure driven flows and three pressure differences.
\label{fig4}} \vspace{-5mm}
\end{center}
\vspace{-5mm}
\end{figure}

When the microactuator buckles into the channel, the interface between two fluids adopts a shape with a peak that propagates through the channel while the amplitude attenuates (Fig.~\ref{fig4}a). The latter phenomenon, which can be treated similar to the attenuation of surface waves within viscous fluids,\cite{Landau1987,Behroozi2004} suggests that the interface amplitude and its location can be obtained more accurately at the beginning of merging the two channels. The flow rate associated with each applied pressure can also be estimated by tracking the location of the amplitude as it attenuates along the channel. These average flow rates were 0.5, 5.8, and 8.4 mm/s, for applied pressures of 250, 800, and 1600 Pa, respectively. The fluid viscosity of solution in the main channel was $1.8 \text{ mPa.s}$ (Vibro Viscometer SV-10) and its density,  measured by weighing 30 ml of fluid using an analytical balance (Sartorius Practum 124), was $1120 \text{ kg}/\text{m}^3$. Knowing the channel's geometry ($H=50 \,\mu \text{m and }w=400 \,\mu \text{m}$), $\mathcal{R}e$ was then calculated ($\mathcal{R}e = $ 0.03, 0.36, and 0.52 for $P_{applied} = 250, \,800, \text{ and }1600 \text{ Pa}$, respectively). 

We then applied the same voltage magnitude at higher frequencies (Fig.~\ref{fig4}b) and observed the microactuator's behavior at different frequencies and different applied pressures. The maximum amplitudes within each voltage cycle were obtained, averaged for a certain frequency, and plotted in Fig.~\ref{fig4}c. When the applied pressure is relatively high, $e.g. \, P>800$ Pa, increasing the actuation frequency slightly improves the maximum amplitude. When the flow rate is very low, higher actuation frequencies, however, reduce the maximum amplitude. This transition appears to occur as the frequency exceeds $f=3.5 \pm 0.5$ Hz. We also note that the amplitude is much higher when the flow rate is slow, as the film deformation can disturb the yellow flow and consequently create a big wave along the channel. 

An increase in the amplitude does not necessarily mean that the actuator undergoes higher postbuckling deformations. In order to estimate the film deformation, we calculate the area under the disturbed fluid interface within the persistence length of the altered fluid flow. Since the fluids are miscible, we do not expect the interface between them to be vertical far downstream;\cite{Ismagilov2000} however, this is a reasonable assumption near the Y--junction -- as the Peclet number is also greater than 500 for all the experiments -- which enables a coarse estimation of the fluid volume moved as a result of the film deformation (Fig.~\ref{fig4}d). The results suggest that when the fluid flow is relatively high, the thin film undergoes more buckling as the actuation frequency increases. When the flow is low, however, the energetic cost for the thin film to move the fluid becomes higher than the cost of the film to buckle into higher, asymmetric modes.\cite{Tavakol2014} On the other hand, a regular flow over the thin film may help it to deform more, similar to the deformation of flexible tubes due to their inner flows.\cite{Grotberg2004,Shapiro1977} We note that the determination of the persistence length significantly affects the estimation of disturbed volume; further investigation is therefore required to verify the trend for other cases, $e.g.$ when $\Delta p \approx 0$.

Based on the 3D patterns of deformation (Fig.~\ref{fig2}b), we assume that the thin film undergoes the first buckling mode when $\Delta p=800 $ Pa or $\Delta p = 1600$ Pa. 
Therefore, the amount of fluid moved by the film at its maximum deformation can be estimated from the volume of a spherical cap $\mathcal{V} = (3a-w)\pi w^2 /3$ where $a = 400\, \mu\text{m}$ is the plate radius, $\mathcal{V}$ is the fluid volume and, $w$ is the maximum buckling deformation of the thin film. On the other hand, the fluid volume at the maximum amplitude within a time frame can also be estimated using a similar method to the one in Fig.~\ref{fig4}d. 
For example, when the flow rate is about $6 \text{ mm/s}$ and $\Delta p = 50 \text{ Pa}$, the maximum amplitude is about $20 \,\mu\text{m}$ for higher frequencies (Fig.~\ref{fig4}c), so the fluid volume within one frame is about $0.3 \text{ nL}$. Using the above equation, the maximum deformation of the thin film is about $15~\mu\text{m}$,  which is reasonable compared to the channel height of 50 $\mu$m.

These microactuators can work in series and parallel for further functionality, e.g., bidirectional flows, or for more efficiently enhancing (pumping) or controlling (valving) different flows within microchannels. Controlling the flow can significantly be improved by fabricating the main channel with a semicircular cross section.\cite{Unger2000} Unlike other types of electrodes, electrical conductivity of fluid electrodes does not decrease by stretching the dielectric film, resulting in a consistent, large actuation. Since we use highly conductive fluid electrodes, the majority of the voltage drop occurs within the thin film. However, if an experiment needs low conductive fluid electrodes or there are particles and cell sensitive to voltage application, a secondary channel coupled with the main channel would shield the fluid any applied voltage. For instance, the voltage applied to one actuator deforms into the secondary channel, which in turn causes the passive microactuator to partially close the main channel without any electricity involved.

We presented a new means to alter and control fluid flow using the buckling deformation of a microactuator that consists of a thin dielectric film actuated by conductive fluid electrodes. The fluid electrodes reduced the critical buckling voltage and significantly enhanced the rate of deformation. The buckling shape of the micro actuator was qualitatively obtained via tracking intensity changes within dyed fluids. We showed that the disturbed interface between two fluids can be used to characterize the interaction between the soft actuator and corresponding fluid, resulting in estimations of fluid flow and the film's deformation. The pressure difference between the main and control channels may affect the buckling shape, while the applied pressure and the pressure-driven flow may considerably alter the film's deformation. These microactuators may play a significant role in the development of microfluidics where either high amplitude or high speed alteration of the fluid's flow rate is desired.

\vspace{-5mm}
\section*{Experimental}\vspace{-4mm}
The thin film was prepared by spin coating polydimethylsiloxane (PDMS) (Dow Corning Sylgard 184) onto a Silicon wafer with the spin speed and time varied to control the film thickness ($h=\mathcal{O}(50 \mu\text{m})$), and then cured at 90 $^{\circ}$C for 5 min. The molds for top and bottom layers of the multilayer microfluidic channel were prepared by patterning the main and control channels on a Silicon wafer via photolithography, and then etching the wafer using Deep Reactive Ion Etching (DRIE). PDMS was cast onto the resulting molds and cured at 90 $^{\circ}$C for 30 min. The thin film was bonded to the top layer using a Corona treater (Laboratory Cronoa Treater, UV Process Supply Inc). The Bottom layer was also exposed to corona plasma for about 1 minute, aligned under microscope, and bonded to the free side of the thin film. 

The fluid electrode solution was prepared by dissolving 20\% NaCl salt into distilled water at 90 $^{\circ}$C. Water soluble dyes were added to the saltwater solution to differentiate the flows in the different channels. The electrical conductivity of all solutions was very high ($>200 \text{ mS/cm}$ measured using Hanna Instruments HI2003 - Edge). The sample was mounted on a \textit{x-y-z} stage, and imaging was performed with a microscopic lens (Navitar Zoom 6000) attached to a Nikon D610 camera from the top view. The main channels were all grounded, while one control channel at a time was connected to the high voltage signal generated using a high voltage amplifier (Trek 20/20C-HS). LabVIEW was used to generate low voltage signals for controlling the high voltage amplifier, and to track the voltage and current via an NI DAQ System (NI cDAQ-9174 with NI 9205 and NI 9269). The fluid in control channels was stationary while the flow in main channels was pressure driven, from a raised reservoir. The pressure difference between the control and the main channels was also applied by relatively elevating or lowering the corresponding reservoirs. 
 
We are grateful to the National Science Foundation (CMMI-1505125) for financial support.

\end{document}